\documentclass[onecolumn,showpacs,nobibnotes,nofootinbib,10pt]{revtex4}
\usepackage{amsmath,amssymb}
\usepackage{epsfig}

\begin{document}
\rightline{BI-TP 2006/42}
\rightline{CU-TP 1168}
\title{Total gluon shadowing due to fluctuation effects}
\author{Misha~Kozlov}
\email{mkozlov@physik.uni-bielefeld.de}
\author{Arif~I.~Shoshi}
\email{shoshi@physik.uni-bielefeld.de}
\affiliation{Fakult{\"a}t f{\"u}r Physik, Universit{\"a}t Bielefeld, D-33501 Bielefeld,
Germany}
\author{Bo-Wen Xiao}
\email{bowen@phys.columbia.edu}
\affiliation{Department of Physics, Columbia University, New York, NY, 10027, USA}

\begin{abstract}
  
  We show a new physical phenomenon expected for the ratio $R_{pA}$ of the
  unintegrated gluon distribution of a nucleus over the unintegrated gluon
  distribution of a proton scaled up by the atomic factor $A^{1/3}$ in the
  fluctuation-dominated (diffusive scaling) region at high energy. We calculate
  the dependence of $R_{pA}$ on the atomic number $A$, the rapidity $Y$ and the
  transverse gluon momentum $k_{\perp}$. We find that $R_{pA}$ exhibits an
  increasing gluon shadowing with growing rapidity, approaching $1/A^{1/3}$ at
  asymptotic rapidities which means total gluon shadowing, due to the effect of
  gluon number fluctuations or Pomeron loops. The increase of $R_{pA}$ with
  rising gluon momentum decreases as the rapidity grows. In contrast, in the
  geometric scaling region where the effect of fluctuations is negligible, the
  ratio $R_{pA}$ shows only partial gluon shadowing in the fixed-coupling case,
  basically independent on the rapidity and the gluon momentum.

\end{abstract}


\maketitle
\section{Introduction}
\label{sec:intro}
There has been a tremendous progress towards understanding gluon number
fluctuation or Pomeron loop effects in the high energy evolution in QCD over
the last two years.  QCD evolution equations have been established, sometimes
called "Pomeron loop
equations"~\cite{Mueller:2005ut,Iancu:2004iy,Iancu:2005nj}, which take into
account fluctuations and a relation between high density QCD and
reaction-diffusion processes in statistical physics has been found which allows
us to obtain universal, analytical results for scattering amplitudes in the
fluctuation-dominated regime at high
energy~\cite{Munier:2003vc,Iancu:2004es,Iancu:2004iy}. The main result , as a
consequence of fluctuations, is the emerge of a new scaling behaviour for the
dipole-hadron (proton or nucleus) scattering amplitude at high
rapidities~\cite{Mueller:2004se,Iancu:2004es}, different from the geometric
scaling behaviour which is the hallmark of the "mean field" evolution equations
(JIMWLK~\cite{Jalilian-Marian:1997jx+X} and
BK~\cite{Balitsky:1995ub+X,Kovchegov:1999yj+X} equations), and which is named
\textit{diffusive scaling} by the authors of Ref.~\cite{Hatta:2006hs}. The
effect of fluctuations on the scattering
amplitude~\cite{Kovner:2005nq,Levin:2005au,Enberg:2005cb,Brunet:2005bz,Munier:2006um,Shoshi:2005pf,Bondarenko:2006rh,Blaizot:2006wp,Iancu:2006jw,Marquet:2005ak,Soyez:2005ha,Kozlov:2006zj,Bondarenko:2006ft},
the diffractive scattering
processes~\cite{Hatta:2006hs,Kovner:2006ge,Shoshi:2006eb,Kozlov:2006cu} and
gluon production in hadronic scattering
processes~\cite{Iancu:2006uc,Kovner:2006wr} has been studied so far. In this
work we focus on the consequences of fluctuations on the ratio $R_{pA}$ of the
unintegrated gluon distribution of a nucleus over the unintegrated gluon
distribution of a proton scaled up by the atomic factor $A^{1/3}$.

The "Pomeron loop equations" describing the evolution of the dipole-hadron
scattering amplitude with increasing rapidity $Y=\ln(1/x)$ are stochastic
equations. In a frame where most of the rapidity is given to the
hadron, the stochastic evolution of the hadron gives rise to different gluon
distributions from one event to another~\cite{Iancu:2004es}. The random variable
in the evolution, the logarithm of the saturation momentum $\rho_s(A,Y)=
\ln(Q^2_s(Y)/k_0^2)$, can therefore vary in an event by event basis. This
variation is characterized by the dispersion $\sigma^2 = \langle \rho_s^2\rangle
- \langle \rho_s\rangle^2$ which rises linearly with rapidity, $\sigma^2(Y) =
D_{\mbox{\scriptsize{dc}}} \bar{\alpha}_s Y$, with $D_{\mbox{\scriptsize{dc}}%
} $ the dispersion coefficient and $\bar{\alpha}_s = \alpha_s N_c/\pi$. For rapidities $Y<Y_{DS}$%
, the dispersion is small $\sigma^2 \ll 1$, meaning that the effect of
fluctuations on the scaling form of the unintegrated gluon distribution can be neglected, while for $Y>Y_{DS}$, where $%
\sigma^2 \gg 1$, fluctuations become important and do change the scaling form of
the unintegrated gluon distribution. We will calculate the gluon distribution of
the proton and the nucleus in the geometric scaling region at $Y<Y_{DS}$ and in
the diffusive scaling region at $Y>Y_{DS}$ which are need in order to study
$R_{pA}$ as a function of the atomic number $A$, the rapidity $Y$ and the transverse
gluon momentum $k_{\perp}$. All calculations which are presented in this work
are valid in the case of a fixed coupling $\alpha_s$. 
 
In the overlapping geometric scaling regime of the proton and the nucleus
$R_{pA}$ has been studied by using the JIMWLK or BK
equations~\cite{Mueller:2003bz,Iancu:2004bx,Baier:2003hr,Kharzeev:2003wz,Jalilian-Marian:2003mf,Blaizot:2004wu,Albacete:2003iq,Baier:2004tj}.
It was found~\cite{Mueller:2003bz,Iancu:2004bx} that $R_{pA}$ scales with $A$
like $A^{1/3(\gamma_c-1)}$ in the fixed-coupling case, rather than being equal
to one ($\gamma_c = 0.6275$) as in the standard pQCD region at very large gluon
momenta. This is partial gluon shadowing due to the anomalous behaviour of the
unintegrated gluon distribution which stems from the BFKL evolution. Partial
gluon shadowing may explain why particle production in heavy ion collisions
scales, roughly, like $N_{part}$~\cite{Kharzeev:2002pc}. Furthermore $R_{pA}$
turns out basically $k_{\perp}^2$ and $Y$ independent in the common geometric
scaling regime. The explanation is that the unintegrated gluon distributions of
the nucleus and of the proton preserve their shapes with rising rapidity in the
geometric scaling regime, yielding thus a constant value for their ratio, as
shown for two different rapidities in Fig.~\ref{GS}(a). We will
derive the known results in the geometric scaling regime in this work in order
to have a direct comparison with the results in the diffusive scaling regime.

We have found that in the overlapping diffusive scaling regime of the proton and
the nucleus $R_{pA}$ basically behaves, for a fixed coupling, as
\begin{equation}
R_{pA} \simeq A^{\frac{1}{3}(\frac{\Delta\rho_s}{2\sigma^2}-1)}\ 
\left[\frac{k_{\perp}^2}{\langle
  Q_s(A,y)\rangle^2}\right]^{\frac{\Delta\rho_s}{\sigma^2}}
\label{R_pA1}
\end{equation}
where $\Delta\rho_s $ denotes the difference between the average saturation
lines of the nucleus and the proton and $\langle Q_s(A,y)\rangle$ is the
average saturation momentum of the nucleus. This ratio shows two features which
are different as compared to the
ratio in the geometric scaling regime: (i) For $k^2_{\perp}$ close to $%
\langle Q_s(A,Y)\rangle^2$, the gluon shadowing characterized by
$A^{\frac{1}{3}(\frac{\Delta\rho_s}{2\sigma^2}-1)}$ is dominated by fluctuations, through
$\sigma^2(Y)$, and depends also on the difference $\Delta\rho_s$. Gluon shadowing
increases as the rapidity increases because of $\sigma^2 = D_{\mbox{\scriptsize{dc}}} \bar{\alpha}_s Y$. At asymptotic rapidity, where $\sigma^2 \to \infty$%
, one obtains \textit{total gluon shadowing}, $R_{pA}=1/A^{1/3}$, which means
that the unintegrated gluon distribution of the nucleus and that of the proton
become the same in the diffusive scaling regime. Total gluon shadowing is an
effect of fluctuations since the fluctuations make the unintegrated gluon
distributions of the nucleus and of the proton flatter and
flatter~\cite{Iancu:2004es} and their ratio closer and closer to $1$ (at fixed
$\Delta\rho_s$) with rising rapidity as illustrated in Fig.\ref{GS}(b). (Total
gluon shadowing is not possible in the geometric scaling regime in the
fixed-coupling case since the shapes of the gluon distributions of the nucleus
and of the proton remain the same with increasing $Y$ giving for their ratio a
value unequal one, see Fig~\ref{GS}(a)). (ii) $R_{pA}$ shows an increase with rising $%
k^2_{\perp}$ (always $R_{pA}< 1$) within the diffusive scaling region. Since the exponent
$\Delta\rho_s/\sigma^2$ decreases with rapidity, the slope of $R_{pA}$ as a
function of $k_{\perp}^2$ becomes smaller with increasing $Y$. The behaviour of
$R_{pA}$ as a function of $k_{\perp}$ with increasing rapidity in the diffusive
scaling regime is shown in Fig.~\ref{R_DS}.

The JIMWLK~\cite{Jalilian-Marian:1997jx+X} or
BK~\cite{Balitsky:1995ub+X,Kovchegov:1999yj+X} equations, with the geometric
scaling being their main consequence, appear to be appropriate for the
understanding of the physics explored at RHIC and HERA experiments. However, it
may be that this isn't the case anymore at LHC energies where fluctuations may
start becoming nonnegligible. Therefore, in addition to the theoretically
interesting results for $R_{pA}$ as a consequence of fluctuations, our result
for $R_{pA}$ may also become relevant in the range of LHC energy. If this is the
case, then the increase of the gluon shadowing and the decrease as a function of
the gluon momentum of $R_{pA}$ with rising rapidity as given by
Eq.~(\ref{R_pA1}) may be viewed as signatures for the onset of fluctuation
effects in the LHC data.

This work is organized as follows: In Sec.~\ref{sec:gen} we show two different
definitions of the unintegrated gluon distributions used in the literature,
define $R_{\mbox{\scriptsize{pA}}}$ and discuss the initial conditions for
the proton and nucleus gluon distributions. In Sec.~\ref{sec:ugd_hadron} we show
the results for the gluon distribution of the proton and the nucleus in the
geometric and diffusive scaling regime. In Sec.~\ref{sec:ratio} we briefly
review the known results for the ratio $R_{pA}$ in the geometric scaling regime,
then calculate and discuss the new results for $R_{\mbox{\scriptsize{pA}}}$ in
the diffusive scaling regime.

\section{Generalities; definitions, initial conditions}
\label{sec:gen}
In this section we show two different definitions for the unintegrated gluon
distribution used in the literature, define the ratio
$R_{\mbox{\scriptsize{pA}}}$ and describe the initial conditions we use for the
unintegrated gluon distribution of the proton and the nucleus.

\subsection{Definition of unintegrated gluon distributions and $R_{
\mbox{\scriptsize{pA}}}$}
\label{sec:ugd_R}
Both definitions of the unintegrated gluon distributions of a hadron (proton or
nucleus) known in the literature are expressed in terms of the forward
scattering amplitude $N\left(x_{\perp},b_{\perp},Y\right)$ of a QCD dipole of
transverse size $x_{\perp}$ with rapidity $Y=\ln 1/x$ scattering off a hadron at
impact parameter $b_{\perp}$. Hereafter we consider the scattering process at a
fixed impact parameter and omit therefore the
$b_{\perp}$-dependence in all the following formulae.

The following definition has been proposed in Ref.~\cite{Braun:2000wr} for
the unintegrated gluon distribution:
\begin{eqnarray}
h_{A}\left( k_{\perp },Y\right) &=&\frac{N_{c}}{\left( 2\pi \right)
^{3}\alpha _{s}}\,\int d^{2}x_{\perp }\,e^{ik_{\perp }\cdot x_{\perp }}\,\nabla
_{x_{\perp }}^{2}\,N\left( x_{\perp },Y\right) ,  \label{k} \\
&=&\frac{N_{c}}{\left( 2\pi \right)^{3}\alpha _{s}}k_{\perp }^{2}\nabla
_{k_{\perp }}^{2}\,\int \frac{d^{2}x_{\perp }}{x_{\perp }^{2}}\, e^{ik_{\perp
}\cdot x_{\perp }}\,N\left( x_{\perp },Y\right) .
\end{eqnarray}
This formula usually appears in cross sections for gluon production in
proton-nucleus collisions~\cite{Baier:2004tj}. The inversion of Eq.~(\ref{k}%
) being
\begin{equation}
N\left( x_{\perp },Y\right) =\frac{\left( 2\pi \right) ^{3}\alpha _{s}}{%
2N_{c}}\,\int \frac{d^{2}k_{\perp }}{\left( 2\pi \right)^{2}k_{\perp }^{2}}%
\ h_{A}\left( k_{\perp },Y\right)\ \left( 2-e^{ik_{\perp }\cdot x_{\perp
}}-e^{-ik_{\perp }\cdot x_{\perp }}\right)
\end{equation}
does show more explicitly the relation between $h_{A}\left( k_{\perp },Y\right)$
and the dipole-nucleus scattering amplitude: The phase factors in the bracket
are due to the four graphs describing the different ways of gluon exchange
between the quark and antiquark of the dipole and the hadron.

The other definition of the unintegrated gluon distribution reads
\begin{equation}
\,\varphi _{A}\left( k_{\perp },Y\right) =\frac{N_{c}}{\left( 2\pi \right)
^{3}\alpha _{s}}\,\int \frac{d^{2}x_{\perp }}{x_{\perp }^{2}}\,e^{ik_{\perp
}\cdot x_{\perp }}\,N\left( x_{\perp },Y\right)  \label{sa}
\end{equation}
and is derived from the non-Abelian Weizsacker-Williams field of a
nucleus~\cite{Kovchegov:1996ty+X,Jalilian-Marian:1996xn+X,Kovchegov:1998bi}. This distribution counts the number of gluons in the
wavefunction of the hadron. The two distributions are related to each other by
\begin{equation}
h_{A}\left( k_{\perp },Y\right) =k_{\perp }^{2}\nabla _{k_{\perp }}^{2}\varphi
_{A}\left( k_{\perp },Y\right) \ . 
\label{eq:re;_h_v}
\end{equation}
The main result of this work, the ratio $R_{pA}$ in the diffusive scaling
regime, turns out to be basically the same for both distributions, as shown in
Sec.~\ref{sec:ratio}. Some more elaborate discussion on the two different
definitions of the unintegrated gluon distribution can be found in
Refs.~\cite{Kharzeev:2003wz,Iancu:2004bx}.

The quantity which we study in this work is the ratio of the
unintegrated gluon distribution of a nucleus over the unintegrated gluon
distribution of a proton scaled up by $A^{1/3}$,
\begin{equation}
R^h_{pA} = \frac{h_{A}\left( k_{\perp },Y\right) }{A^{\frac{1}{3}%
}\ h_{p}\left( k_{\perp },Y\right) }
\quad \quad {\mbox{and}} \quad \quad
R^{\varphi}_{pA}= \frac{\varphi_{A}\left(
k_{\perp },Y\right) }{A^{\frac{1}{3}}\ \varphi_{p}\left( k_{\perp
},Y\right) }\ ,
\label{R_pA}
\end{equation}
which is a measure of the ratio of the number of particles produces in a
proton-nucleus collisions and the corresponding number in proton-proton
collisions scaled up by the number of collisions~\cite{Kharzeev:2003wz,Iancu:2004bx,Iancu:2006uc,Jalilian-Marian:2003mf,Blaizot:2004wu,Baier:2004tj}.

\subsection{Initial condition in the case of a nucleus}
\label{sec:init_cond}
The evolution of the unintegrated gluon distributions is
given by the QCD evolution equations. However, one has to know the initial
condition for the unintegrated gluon distribution to start with in the
evolution equations. In the case of a nucleus, as an initial condition we
use the McLerran-Venugopalan model~\cite{McLerran:1993ka+X} at fixed $b_{\perp}$ and $Y=0$,
\begin{equation}
N_{MV}\left(x_{\perp },Y=0\right) = 1-\exp \left(-\frac{x_{\perp }^{2}\
Q^2_s(A)}{4}\right) ,  \label{mv}
\end{equation}%
with the saturation momentum given by~\cite{McLerran:1993ka+X}

\begin{equation}
Q^2_s(A) = \frac{2\pi ^{2}\alpha _{s}}{N_{c}} \rho T\left(b\right)xG\left(x,%
\frac{1}{x_{\perp }^{2}}\right)
\end{equation}
where $\rho$ is the nuclear density, $T\left(b\right)=2\sqrt{R_{A}^2-b^2}$ is
the profile function and $xG\left(x,1/ x_{\perp }^{2}\right)$ is the gluon
distribution in the nucleon which at $Y=0$ can be written as $xG\left(x=1,1/
  x_{\perp }^{2}\right) \propto \alpha_s \ln (1/x_{\perp}^2 \Lambda_{QCD}^2)$.
Since we are interested throughout this work in the region where $1/x_{\perp}$
is not much larger than the saturation scale $Q^2_s(A)$, we neglect the
$x_{\perp}$ by replacing it by $1/Q^2_s(A)$ in $xG(x,1/x_{\perp}^2)$. Noting
also that $\rho\,T(b) \propto Q_0^2 A^{1/3}$ at fixed impact parameter, one gets
for the $\alpha_s$ and $A$ dependence
\begin{equation}
Q_s^{2}(A) \propto
Q_0^2\, \alpha_s^2 A^{\frac{1}{3}}\, \ln(\alpha_s^2 A^{1/3})
\end{equation}
which has also been used in
Refs.~\cite{Mueller:2003bz,Iancu:2006uc}. We shall always assume that $\alpha_s^2 A^{\frac{1}{3}}\gg 1$ in order to have non-trivial
nuclear effects. 

Using the Mellin transform of the scattering amplitude,
\begin{equation}
N_{MV}\left( \gamma \right) =\int_{0}^{\infty }duu^{-\gamma -1}\left[ 1-\exp
\left( -u\right) \right] =-\Gamma \left( -\gamma \right) \text{ }%
\mbox{ with
}\text{ }0<\gamma <1
\end{equation}%
where $u=\frac{x_{\perp }^{2}Q_s^2(A)}{4}$, one can write Eq.~(\ref{mv})
also in the form
\begin{equation}
N_{MV}\left(x_{\perp},Y =0\right) = \int \frac{ d\gamma }{2\pi i}%
N_{MV}\left( \gamma \right) \exp \left[ \gamma \ln \left( \frac{x_{\perp
}^{2}Q_s^2(A)}{4}\right) \right] \ ,
\label{mv1}
\end{equation}
which together with Eq.~(\ref{k}) and Eq.~(\ref{sa}) allows us to get the
unintegrated gluon distribution of a nucleus at $Y=0$,
\begin{eqnarray}
h_{A}\left( k_{\perp },Y=0\right) &=&\frac{N_{c}}{2\pi ^{2}\alpha _{s}}%
\left( \frac{k_{\perp }^{2}}{Q_s^{2}(A)}\right) \exp \left( -\frac{k_{\perp
}^{2}}{Q_s^{2}(A)}\right) \ , \label{h0}\\
\varphi _{A}\left( k_{\perp },Y=0\right) &=&\frac{N_{c}}{8\pi ^{2}\alpha _{s}%
}\Gamma \left( 0,\frac{k_{\perp }^{2}}{Q_s^{2}(A)}\right) , \label{v0}
\end{eqnarray}
where $\Gamma \left( 0,\frac{k_{\perp }^{2}}{Q_s^{2}(A)}\right)\simeq
\ln\left(\frac{k_{\perp }^{2}}{Q_s^{2}(A)}\right)-\gamma_E+\frac{k_{\perp
}^{2}}{Q_s^{2}(A)}+...$ is the incomplete Gamma function.

Let us also construct the expressions for unintegrated gluon
distributions which take into account BFKL evolution since they
turn out to be useful in the next sections. This is easily done by
starting with the BFKL evolved scattering amplitude in terms of
the Mellin transform
\begin{equation}
N_{MV}\left( x_{\perp },Y\right) =\int_{c-i\infty }^{c+i\infty }\frac{%
d\gamma }{2\pi i}N_{MV}\left( \gamma \right) \exp \left[ \overline{\alpha }%
_{s}Y\chi (\gamma )+\gamma \ln \left( \frac{x_{\perp }^{2}Q^2_s(A)}{4}%
\right) \right]  \label{bfkl}
\end{equation}
where $0<c<1$, $\overline{\alpha }_{s}=\frac{\alpha _{s}N_{c}}{\pi }$ and $\chi
(\gamma )=2\psi (1)-\psi (\gamma )-\psi (1-\gamma )$ with $\psi (\gamma )$ a
polygamma function. With Eq.(\ref{bfkl}) inserted into Eq.(\ref{k}) and
Eq.(\ref{sa}) and the relation
\begin{equation}
\,\int \frac{d^{2}x_{\perp }}{x_{\perp }^{2}}e^{ik_{\perp }\cdot x_{\perp
}}\,\left( \frac{x_{\perp }^{2}Q_s^{2}(A)}{4}\right) ^{\gamma }=\pi \frac{%
\Gamma \left( \gamma \right) }{\Gamma \left( -\gamma +1\right) }\left( \frac{%
k_{\perp }^{2}}{Q_s^{2}(A)}\right) ^{-\gamma }
\end{equation}
one obtains the Mellin representation of the unintegrated gluon
distributions which contain rapidity evolution:
\begin{eqnarray}
h_{A}\left( k_{\perp },Y\right) &=&\frac{N_c}{2 \pi^2\,\alpha_s}\ \int_{c-i\infty }^{c+i\infty }\frac{%
d\gamma }{2\pi i}\ \Gamma \left( \gamma +1\right)\ \exp \left[ \overline{%
\alpha }_{s}Y\chi (\gamma )-\gamma \ln \left( \frac{k_{\perp }^{2}}{%
Q_s^{2}(A)}\right) \right] ;  \label{h} \\
\varphi_{A}\left( k_{\perp },Y\right) &=& \frac{N_c}{8 \pi^2\,\alpha_s}\ \int_{c-i\infty }^{c+i\infty }%
\frac{d\gamma }{2\pi i}\ \frac{\Gamma \left( \gamma \right) }{\gamma }\ \exp %
\left[ \overline{\alpha }_{s}Y\chi (\gamma )-\gamma \ln \left( \frac{%
k_{\perp }^{2}}{Q_s^{2}(A)}\right) \right] .  \label{h2}
\end{eqnarray}
In the limit of $Y=0$ Eq.~(\ref{bfkl}), Eq.~(\ref{h}) and
Eq.~(\ref{h2}) reduce, of course, to Eq.(\ref{mv1}), Eq.(\ref{h0})
and Eq.(\ref{v0}) respectively.

\subsection{Initial condition in the case of a proton}
\label{sec:prot_IC}
We use two different ways to describe the initial gluon distribution of the
proton: (i) In the main body of this work we view the proton as a single color
dipole (see also Ref.~\cite{Mueller:2003bz}) and (ii) in
Appendix~\ref{MV_proton} we view the proton as a composite object and use the
McLerran-Venugopalan model to describe its initial conditions (see also
Ref.~\cite{Iancu:2004bx}). We will show that both point of views exhibit equally
well the effect of fluctuations on $R_{pA}$ which is the main focus of this
work.

The scattering amplitude for a dipole of size $x_{\perp}$ scattering off a dipole
of size $x^{\prime}_{\perp}$ at relative rapidity $Y$ reads~\cite{Mueller:2002zm}
\begin{equation}
N(x^{\prime}_{\perp},x_{\perp},Y) = \pi\, \alpha_s^2\, x^{\prime\,2}_{\perp} \int_{c-i\infty}^{c+i\infty}
\frac{d\gamma}{2\pi i}\,\frac{1}{\gamma^2(1-\gamma)^2} \
  \exp\left[\bar{\alpha}_s \chi(\lambda) Y - \gamma\,
    \ln\frac{x^{\prime\,2}_{\perp}}{x^2_{\perp}}\right]
\end{equation}
where $0 <c < 1$. When $Y=0$ the above expression reduces to the dipole-dipole
scattering amplitude at the two gluon exchange level
\begin{equation}
N(x^{\prime}_{\perp},x_{\perp},Y=0) = 2 \pi\, \alpha_s^2\, r^2_{<}\
\left(1+\ln\frac{r_{>}}{r_{<}}\right)
\end{equation}
with $r_{<}$ being the smaller of $x_{\perp}$, $x^{\prime}_{\perp}$ and $r_{>}$
the larger of $x_{\perp}$, $x^{\prime}_{\perp}$.

Following the same steps as in the case of a nucleus, one obtains for the
unintegrated gluon distributions of a proton which take into account BFKL
evolution
\begin{eqnarray}
h_{p}\left( k_{\perp },Y\right) &=&\frac{N_c}{2 \pi^2\,\alpha_s}\
\pi\,\alpha^2_s\,x^{\prime\,2}_{\perp}\ \int_{c-i\infty }^{c+i\infty }\frac{%
d\gamma }{2\pi i}\ \frac{4^{\gamma}}{(1-\gamma)^2}\,\frac{\Gamma(\gamma)}{\Gamma(1-\gamma)}\ \exp \left[ \overline{%
\alpha }_{s}Y\chi (\gamma )-\gamma \ln \left(k_{\perp }^{2} x^{\prime\,2}_{\perp}\right) \right] ;  \label{h_p} \\
\varphi_{p}\left( k_{\perp },Y\right) & = &\frac{N_c}{8 \pi^2\,\alpha_s}\
\pi\,\alpha^2_s\,x^{\prime\,2}_{\perp}\ \int_{c-i\infty }^{c+i\infty }%
\frac{d\gamma }{2\pi i}\ \frac{4^{\gamma}}{\gamma^2(1-\gamma)^2}\,\frac{\Gamma(\gamma)}{\Gamma(1-\gamma)}\ \exp %
\left[ \overline{\alpha }_{s}Y\chi (\gamma )-\gamma \ln \left(k_{\perp }^{2}
    x^{\prime\,2}_{\perp}\right) \right] \ .  \label{h2_p}
\end{eqnarray}
The Eqs.~(\ref{h}), (\ref{h2}) and Eqs.~(\ref{h_p}), (\ref{h2_p})
have the form of the formulae one has started with in
Ref.~\cite{Mueller:2002zm} and
Ref.~\cite{Mueller:2004se,Iancu:2004es} to study the geometric
scaling region and the diffusive scaling region. These equations
tell us what the effect of the different inital conditions is
according to
Ref.~\cite{Mueller:2002zm,Mueller:2004se,Iancu:2004es}: Different
initial conditions for the proton as compared with the ones for
the nucleus do lead to different saturation momenta, $Q_s(p,Y)$
unequal $Q_s(A,Y)$. The saturation momentum does slightly depend
also on the definition for the unintegrated gluon distribution,
$h_{p,A}(k_{\perp},Y)$ and $\varphi_{p,A}(k_{\perp},Y)$. In the
next section we do show the exact expressions of the saturation
momentum of the proton and the nucleus.


\section{Unintegrated gluon distribution of a hadron}
\label{sec:ugd_hadron}
\begin{figure}[htb]
\setlength{\unitlength}{1.cm}
\par
\begin{center}
\epsfig{file=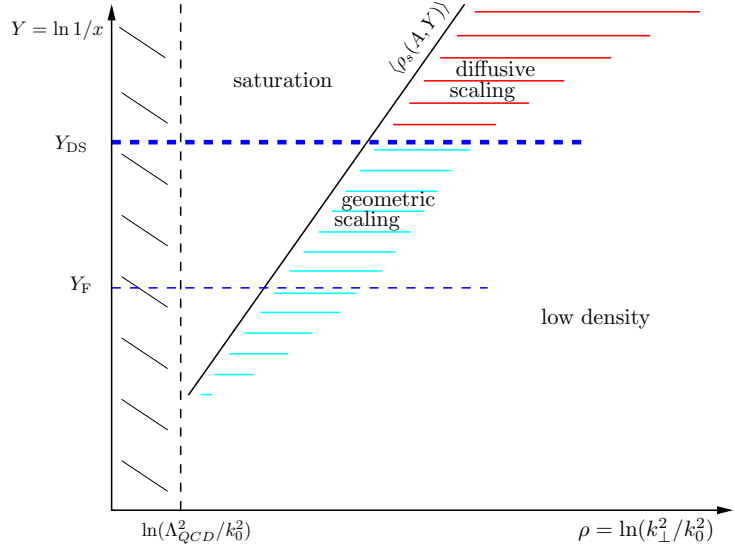, width=10cm}
\end{center}
\caption{The phase diagram of the unintegrated gluon distribution of a proton 
  (or nucleus) in the kinematical plane $Y-\ln(k^2_{\perp}/k_0^2)$. The high
  density region (saturation region) at small gluon momenta is on the left hand
  side of the saturation line $\langle \rho_s(A,Y) \rangle$, the low density
  region at very high gluon momenta is one the right hand side of shadowed
  area. The intermediate region separated by $Y_{\mbox{\scriptsize{DS}}} $
  divides the geometric scaling regime from the diffusive scaling regime.}
\label{had_wf}
\end{figure}

In this section we focus on the unintegrated gluon distribution of a highly
evolved hadron (nucleus or proton) in the geometric and diffusive scaling
region. To explain the relevant physics in these two regions let us
look at the phase diagram of the hadron in the high-energy limit shown in the $%
Y-\rho $ plane in Fig.~\ref{had_wf}.  Here $Y=\ln (1/x)$ is the rapidity of the
hadron and $\rho =\ln (k_{\perp }^{2}/k_{0}^{2})$ is the logarithm of the
transverse momentum of gluons inside the hadron ($k_{0}^{2}$ is a fixed
reference scale). The straight line denoted by $\langle \rho _{s}(A,y)\rangle $
represents the average saturation line, $\langle \rho _{s}(A,Y)\rangle =\langle
\ln (Q_{s}(A,Y)/k_{0}^{2})\rangle $, with the saturation momentum $Q_{s}(A,y)$
depending on $Y$ and the atomic number $A$. To the left of the saturation line,
$\rho < \langle \rho_s(A,Y) \rangle$, the gluon density is of order $1/\alpha
_{s}$, or, the dipole-hadron scattering amplitude is at the unitarity limit
$\langle N\rangle \approx 1$. In this region, called "saturation
region", the nonlinear QCD evolution becomes important. For
$\rho \gg \langle \rho_s(A,Y)\rangle$, the gluon density is low and the standard
'double-logarithmic approximation' is applicable. In the low density regime
neither saturation nor fluctuation effects are important and the unintegrated
gluon distribution behaves as $1/k_{\perp }^{2}$ at very large $k_{\perp}$ (the amplitude shows color
transparency, $\langle N(r_{\perp })\rangle \propto r_{\perp }^{2}$).  The most
interesting region for us in this work is the transition region between the
saturation and the low density regime where $\rho$ is not much larger than
$\langle \rho_s(A,Y)\rangle$ (see Fig.~\ref{had_wf}).  There are two different
regimes within the transition region which are separated by the rapidity scale
$Y_{\mbox{\scriptsize{DS}}}$, the \textit{geometric scaling
  regime}~\cite{Iancu:2002tr,Mueller:2002zm,Munier:2003vc} and the
\textit{diffusive scaling regime}~\cite{Mueller:2004se,Iancu:2004es}, in which
the dynamics of the QCD evolution is different.


The QCD evolution of the dipole-hadron scattering amplitude with increasing
rapidity $Y=\ln(1/x)$ is stochastic~\cite{Iancu:2004es,Mueller:2004se} meaning
that the scattering amplitude can fluctuate from event to event. Equivalently,
in a frame where most of the rapidity is given to the hadron, the stochastic
evolution of the hadron gives rise to different gluon distributions from one
event to another~\cite{Iancu:2004es}. Therefore the random variable,
$\rho_s(A,Y) = \ln(Q^2_s(A,Y)/k_0^2)$, can vary from one event to another. Its
variation is characterized by the dispersion $\sigma^2 = \langle
\rho_s^2\rangle - \langle \rho_s\rangle^2$ which rises linearly with rapidity,
$\sigma^2(Y) =
D_{\mbox{\scriptsize{dc}}} \bar{\alpha}_s Y$, with $D_{\mbox{\scriptsize{dc}}%
} $ the dispersion coefficient. For rapidities $Y<Y_{DS}$%
, the dispersion is small $\sigma^2 \ll 1$, meaning that the effect of
fluctuations on the scaling form of the unintegrated gluon distribution can be neglected, while for $Y>Y_{DS}$, where $%
\sigma^2 \gg 1$, fluctuations become important and do change the scaling form of
the unintegrated gluon distribution as we show below (see also
Ref.~\cite{Hatta:2006hs,Iancu:2006uc} for more discussions on the phase
diagram).
%
%
\subsection{Geometric scaling regime}
\label{sec:gsr}
In the geometric scaling regime at $Y\ll Y_{F}$ ($Y_F$ will be fixed below, see
also Fig.~\ref{had_wf}) the QCD evolution is described by the Kovchegov
equation~\cite{Kovchegov:1999yj+X}, or, equivalently, by the BFKL equation in
the presence of a saturation
boundary~\cite{Mueller:2002zm,Munier:2003vc}. Starting with Eq.~(\ref{h}) and
following the calculations in Refs.~\cite{Mueller:2002zm,Munier:2003vc} one
finds for the unintegrated gluon distribution of a nucleus in the case of a
fixed coupling
\begin{equation}
h_{A}\left(k_{\perp },Y\right) =h_{A}^{\max }\ \gamma _{c}\ \left( \frac{%
k_{\perp }^{2}}{Q_{s}^{2}(A,Y)}\right) ^{-\gamma _{c}}\left[ \ln \left(
\frac{k_{\perp }^{2}}{Q_{s}^{2}(A,Y)}\right) +\frac{1}{\gamma _{c}}\right]
\label{eq:h}
\end{equation}
which is valid in the geometric scaling regime
\begin{equation}
Q_s^{2}(A) \leq k^2_{\perp} \leq Q_s^{2}(A)\,\exp\left[\sqrt{4
\chi^{^{\prime\prime}}(\gamma_c) \bar{\alpha}_s} Y\right] \ .
\end{equation}
and scales as a function of the dimensionless variable $k_{\perp}^2/Q_s^2(A,Y)$. The observation of a similar scaling behaviour at the HERA data for DIS at $%
x\leq 0.01$ and any momentum~\cite{Stasto:2000er} supports the theoretical
findings. In the above equation $Q_s^{2}(A,Y)$ denotes the saturation momentum
of the nucleus
\begin{equation}
Q_{s}^{2}(A,Y) = c^h_{A}\,Q_s^{2}(A)\,\frac{\exp[\frac{\chi (\gamma _{c})}{%
\gamma _{c}} \overline{\alpha }_{s}Y]}{\left[ 2\overline{\alpha }_{s}Y\chi
^{\prime \prime }(\gamma _{c})\right] ^{\frac{3}{2\gamma _{c}}}} \ ,
\label{eq:q_S_YC}
\end{equation}
where $c_A$ is a constant and the value of the anomalous
dimension, $\gamma_c = 0.6275$, which comes from the BFKL dynamics in the
presence of saturation, is fixed by $\chi^{\prime}(\gamma_c) =
\frac{\chi(\gamma_c)}{\gamma_c}$. In Eq.~(\ref{eq:h}) $h_A^{\mbox{max}} = \mathcal{O}(1/\alpha_s)$ is the maximum of  $%
h_A(k_{\perp},Y)$ which happens when $k^2_{\perp} = Q^2_s(A,Y)$.

The unintegrated gluon distribution $\varphi_A(k_{\perp},Y)$, defined in  Eq.~(%
\ref{sa}), has in the geometric scaling regime exactly the same form as
$h_A(k_{\perp},Y)$ given in Eq.~(\ref{eq:h}). The only difference, steaming from
the different initial conditions at $Y=0$, see eq.~(\ref{h}) and Eq.~(\ref{h2}),
is a slightly different constant $c^{\varphi}_A$ in the saturation momentum.

To get the parametrical estimates for the rapidities $Y_F$ and $Y_{DS}$, it is
convenient to express the above results in the geometric scaling regime in
terms of the $\rho = \ln k^2_{\perp}/k^2_0$ variable,
\begin{equation}
h_{A}\left(k_{\perp },Y\right) = h_A^{\max }\ \gamma _{c}\ e^{-\gamma_c
(\rho-\rho_s(A,Y))}\ \left(\rho-\rho_s(A,Y) + \frac{1}{\gamma_c}\right)
\label{eq:h1}
\end{equation}
with the region of validity
\begin{equation}
0 \leq \rho-\rho_s(A,Y) \leq \sqrt{4 \chi^{^{\prime\prime}}(\gamma_c) \bar{%
\alpha}_s Y}
\label{eq:dr_gs}
\end{equation}
and
\begin{equation}
\rho_s(A,Y) = \ln \frac{c^h_A Q^{2}_s(A)}{k_0^2} + \frac{\chi (\gamma _{c})}{%
\gamma _{c}}\, \overline{\alpha }_{s}\,Y - \frac{3}{2\gamma _{c}} \ \ln\left[ 2\overline{\alpha }_{s}Y\chi
^{\prime \prime }(\gamma _{c})\right] \ .
\label{eq:rho_S}
\end{equation}

Gluon number fluctuations (Pomeron loops)~\cite{Mueller:2004se,Iancu:2004es} do
change the results (\ref{eq:h1}) and (\ref{eq:rho_S}) which emerge from the
Kovchegov equation as the rapidity increases. These results remain valid so
long as the gluon occupancy in the nucleus, $n(k_{\perp},Y) \propto
h_A(k_{\perp},Y)/\alpha_s$, is much larger than one in the geometric scaling
regime. When $n(k_{\perp},Y) \approx O(1)$ (or the scattering amplitude $N(r,Y)
\approx \alpha_s^2 n(k\sim 1/r,Y)$) the discreteness of the gluons has to be
taken into account, allowing the occupancy to go below one only by becoming
zero~\cite{Mueller:2004se,Iancu:2004es}. The gluon occupancy becomes of order
one at the largest $k_{\perp}$ within the geometric scaling regime, or
$h_A(k_{\perp},Y) \approx \alpha_s$, when $\rho-\rho_s \approx
\frac{1}{\gamma_c}\ \ln(1/\alpha^2_s)$. According to Eq.(\ref{eq:dr_gs}), this
happens at the rapidity
\begin{equation}
Y_{\mbox{\scriptsize{F}}} \propto \frac{1}{\alpha_s} \ \ln^2(1/\alpha^2_s) \
.  
\label{eq:Y_F}
\end{equation}
For $Y_{\mbox{\scriptsize{F}}} \leq Y$ fluctuations do change somewhat
$\rho_s(A,Y)$ given in Eq.~(\ref{eq:rho_S})
to~\cite{Mueller:2004se,Iancu:2004es}
\begin{equation}
\langle \rho_s(A,Y) \rangle = \ln \frac{c^h_A\,Q^{2}_s(A)}{k_0^2} +
\left(\frac{\chi (\gamma _{c})}{\gamma _{c}} - \frac{\pi^2
\gamma_c \chi^{\prime \prime}(\gamma_c)}{2
\ln^2(1/\alpha_s^2)}\right)\ \bar{\alpha}_s Y \ .
\label{eq:rho_S_mod}
\end{equation}
Fluctuations, however, do not change the shape of the unintegrated
gluon distribution in Eq.~(\ref{eq:h1}) in the region
$Y_{\mbox{\scriptsize{F}}} \leq Y \ll Y_{DS}$ because of the
following reason: Starting with a unique initial gluon
distribution at $Y=0$, the stochastic evolution generates an
ensemble of them at rapidity Y, where each of the individual gluon
distributions has the shape given by (\ref{eq:h1}), but the
individual gluon distributions may differ from each other by
translation ($\rho_s(A,Y)$ is random). Based on the relation
between QCD evolution and reaction-diffusion models in statistical
physics~\cite{Iancu:2004es}, the fluctuations are taken into
account by averaging over all individual distributions,
\begin{equation}
\langle h_A(\rho-\rho_s(A,Y))\rangle = \int d\rho_s\ h_A(\rho-\rho_s(A,Y)) \
P(\rho_s-\langle\rho_s\rangle) \ ,
\label{av_gd}
\end{equation}
where the probability distribution of $\rho_s(A,Y)$ is argued to have a
Gaussian form~\cite{Marquet:2006xm},
\begin{equation}
P(\rho _{s})\simeq \frac{1}{\sqrt{2\pi \sigma ^{2}}}\exp \left[ -\frac{%
\left( \rho _{s}-\langle \rho _{s}\rangle \right) ^{2}}{2\sigma ^{2}}\right]
\quad \mbox{for} \quad \rho-\rho_s(A,Y) \ll \gamma_c^2 \sigma^2 \ ,
\label{proba_gauss}
\end{equation}
with the variance
\begin{equation}
\sigma^2 = \langle \rho_s(A,Y)^2 \rangle- \langle \rho_s(A,Y) \rangle^2 = D_{%
\mbox{\scriptsize{dc}}} \bar{\alpha}_s Y 
\label{sigma_v}
\end{equation}
and the diffusion coefficient at asymptotic energies given
by~\cite{Mueller:2004se,Iancu:2004es}
\begin{equation}
D_{\mbox{\scriptsize{dc}}} = \frac{\pi ^{4}\gamma
_{c}\chi^{\prime \prime }(\gamma _{c})}{3\log ^{3}(1/\alpha _{s}^{2})} \ .
\end{equation}
Now, if the variance is small, $\sigma^2 \ll 1$, which is the case as long as the
rapidity is much smaller than  
\begin{equation}
Y_{DS} \simeq \frac{1}{D_{\mbox{\scriptsize{dc}}}\,\bar{\alpha}_s} 
 \ ,
\end{equation}
then the Gaussian distribution in Eq.(\ref{proba_gauss}) is strongly peaked and
therefore the averaged gluon distribution nearly preserves the form of an
individual distribution,
\begin{equation}
\langle h_A(\rho-\rho_s(A,Y)) \rangle \approx h_A(\rho -\langle
\rho_s(A,Y)\rangle) \
\end{equation}
and shows geometric scaling, at least in the window $0 \leq \rho-\rho_s(A,Y) \leq \ln(1/\alpha_s^2)/\gamma_c$. So, at $Y\leq Y_{F}$ the unintegrated gluon
distribution in the geometric scaling regime (see Fig.~\ref{had_wf}) is given by (\ref{eq:h1}) with the
saturation momentum given by (\ref{eq:rho_S}) while at $Y_{F} \leq Y \leq
Y_{DS}$ the unintegrated gluon distribution remains approximately the same while
the saturation momentum changes to the one given in Eq.~(\ref{eq:rho_S_mod}).

The only change in the case of a proton as compared to that of the
nucleus in the geometric scaling regime is the saturation
momentum, $\langle\rho_s(A,Y)\rangle \to \langle \rho_s(p,Y)
\rangle$, which is done by replacing $c^{h,\varphi}_A\,Q^2_s(A)$ in the
expressions for $\langle \rho_s(A,Y)\rangle$ by $c^{h,\varphi}_p\,Q^2_s(p)$,
where $c^{h,\varphi}_p$ is a constant and $Q^2_s(p)\,x^{\prime\,2}_{\perp} =
[\alpha_s \sqrt{\ln(1/\alpha_s)}]^{2/\gamma_c}$. The saturation
momenta of the proton and the nucleus are different due to
different initial conditions which we have explained in
Sect.~\ref{sec:init_cond} and Sect.~\ref{sec:prot_IC}. The
difference between them is
\begin{eqnarray}
\Delta \rho_s \equiv
 \langle \rho_s(A,Y)\rangle -  \langle \rho_s(p,Y) \rangle
 &=& \ln \frac{\langle Q_s(A,Y)\rangle^2}{\langle Q_s(p,Y)\rangle^2} \nonumber \\
 &=& \ln \frac{c\, c^{h,\varphi}_A\,Q_0^2\,x^{\prime\,2}_{\perp}\, \alpha_s^2 A^{1/3}\, \ln(\alpha_s^2 A^{1/3})}{c^{h,\varphi}_p\,[\alpha_s\, \sqrt{\ln(1/\alpha_s)}]^{2/\gamma_c}}
\label{del_rho}
\end{eqnarray}
with $c$ a constant. A different $\Delta \rho_s$ is obtained when the proton is
viewed as a composite object, instead of a single dipole, and is described by
the McLerran-Vnugopalan model, as done in Appendix~\ref{MV_proton}.

\subsection{Diffusive scaling regime}
\label{se:dsr_ugd}
The diffusive scaling regime sets in when the variance is large, $\sigma^2 \gg
1$, which happens at $Y \gg Y_{DS}$. In the diffusive scaling regime the
fluctuations are very important and the unintegrated gluon distribution
changes therefore a lot from one event to another, leading to an average unintegrated gluon
distribution $\langle h_A(\rho-\rho_s(A,Y)) \rangle$ in Eq.(\ref{av_gd}) which
is quite different as compared to the event-by-event gluon distribution
$h_A(\rho-\rho_s(A,Y))$ as we show below. 

We use for the event-by-event unintegrated gluon distributions
\begin{equation}
h_{A}(\rho -\rho _{s}(A,Y))=%
\begin{cases}
h^{max}_A\,\exp \left[ -\left( \rho_{s}(A,Y)-\rho \right) \right] & \quad \mbox{for}
\quad \rho < \rho
_{s}(A,Y), \\
h^{max}_A\,\exp \left[ -\gamma _{c}(\rho -\rho_{s}(A,Y))\right] & \quad \mbox{for}
\quad \rho
>\rho_{s}(A,Y)%
\end{cases}%
;  \label{eq:front}
\end{equation}%
\begin{equation}
\varphi_{A}(\rho -\rho_{s}(A,Y))=%
\begin{cases}
\varphi^{max}_A\,\left[ \frac{1}{2}\left( \rho_{s}(A,Y)-\rho \right) +1\right] &\quad \mbox{for}
\quad
\rho < \rho_{s}(A,Y), \\
\varphi^{max}_A\,\exp \left[ -\gamma _{c}(\rho -\rho_{s}(A,Y))\right] & \quad \mbox{for}
\quad \rho
>\rho _{s}(A,Y)  \label{varphi_k}
\end{cases}%
,
\end{equation}
In the above equations we have used for $\rho > \rho_s(A,Y)$ the leading
contributions valid in the geometric scaling regime (see Sect.~\ref{sec:gsr})
while for $\rho<\rho_s(A,Y)$ (saturation regime) the expressions which come from
solving the Kovchegov equation (see Appendix~\ref{sec:ap}
and~\cite{Kovchegov:1999yj+X})~\footnote{The McLerran-Venugopalan model also gives
  the same expressions as a function of rapidity for $\rho <
  \rho_s(A,Y)$.}. There is a difference between the $h_A$ and $\varphi_A$
distributions in the saturation region coming from their different definitions
in (\ref{k}) and (\ref{sa}). (We have used for simplicity
the geometric scaling result also in the low density regime, see Fig.~\ref%
{had_wf}, where both distributions behave the same way, like $1/k^2_{\perp}$%
, since the main result in the diffusive scaling regime is insensitive to the
low density region.)

It is now easy to show that the average unintegrated gluon
distributions are roughly described by Gaussians,
\begin{eqnarray}
\langle h_{A}(\rho -\rho _{s}(A,Y))\rangle &\simeq & \frac{h^{max}_A}{\sqrt{2 \pi
\sigma^2}} \ \left(\frac{1+\gamma_c}{\gamma_c}\right)\ \exp\left[-\frac{%
(\rho-\langle\rho_s(A,Y)\rangle)^2}{2 \sigma^2}\right] \ ,
\label{eq:h_a_diff} \\
\langle\varphi_{A}(\rho -\rho _{s}(A,Y))\rangle &\simeq & \frac{%
{\varphi}^{max}_A\,\sigma^{3}}{2\sqrt{2\pi }\left[\rho-\langle\rho_s(A,Y)\rangle%
\right]^2}\exp \left[ -\frac{[\rho-\langle\rho_s(A,Y)\rangle]^2}{2\sigma ^{2}%
}\right] \ ,  \label{eq:phi_a_diff}
\end{eqnarray}
and are valid in the diffusive scaling regime,
\begin{equation}
\sigma \ll \rho - \langle\rho_s(A,Y)\rangle \ll \gamma_c\,\sigma^2 \ .
\label{eq:dif_scal_reg}
\end{equation}
Note that in the diffusive scaling regime the unintegrated gluon distribution
scales as a function of the dimensionless variable $(\rho-\langle
\rho_s\rangle)/\sigma(Y)$ which is different as compared to the geometric
scaling $(\rho-\langle \rho_s\rangle)$.  The diffusive scaling extends up to
very large values of gluon momenta $k^2_{\perp}$ since the window $\rho-\langle
\rho_s\rangle \ll \sigma^2$ increases with rapidity, see Eq.(\ref{sigma_v}).
The same result as the one in Eq.~(\ref{eq:h_a_diff}) was obtained in
Ref.~\cite{Iancu:2006uc}, although in a different way, and was shown to give the
cross section for gluon production in the diffusive scaling regime.

The geometric (diffusive) scaling window of a proton as compared
to the geometric (diffusive) scaling window of a nucleus at high
rapidity is shifted to lower momenta due to the smaller saturation
momentum of the proton (see (\ref{del_rho})). The proton and the
nucleus have a common geometric scaling region for rapidities
$Y<Y_{DS}$. The common diffusive scaling for the proton and the
nucleus at $Y \geq Y_{\mbox{\scriptsize{DS}}}^{pA}$ sets in when
the diffusive scaling window of the proton overlaps with the one
of the nucleus,
\begin{equation}
\sigma^2 = D_{\mbox{\scriptsize{dc}}} \bar{\alpha}_s Y \ \gg \ \sigma + \Delta
\rho_s
\label{eq:ods}
\end{equation}
which determines
\begin{equation}
Y^{pA}_{\mbox{\scriptsize{DS}}} \ \propto \
\begin{cases}
\frac{1}{D_{\mbox{\scriptsize{dc}}} \bar{\alpha}_s}\ (\Delta \rho_s)^2 \quad & \mbox{for}\quad
\sigma > \Delta \rho_s \\
\frac{1}{D_{\mbox{\scriptsize{dc}}} \bar{\alpha}_s}\ \Delta \rho_s \quad & \mbox{for}\quad
\sigma < \Delta \rho_s \ .
%
\end{cases}%
\end{equation}
%


\section{The ratio $R_{pA}$ in the geometric and diffusive scaling regime}
\label{sec:ratio}
\begin{figure}[htb]
\setlength{\unitlength}{1.cm}
\par
\begin{center}
\epsfig{file=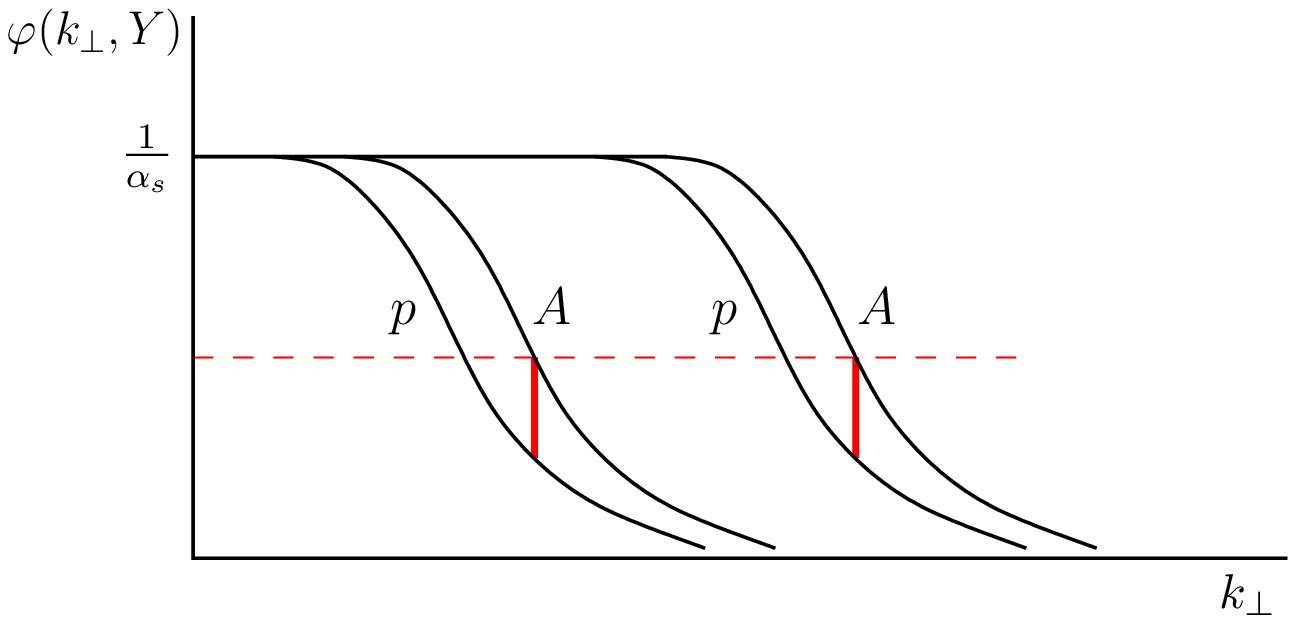, width=8.5cm} \hfill
\epsfig{file=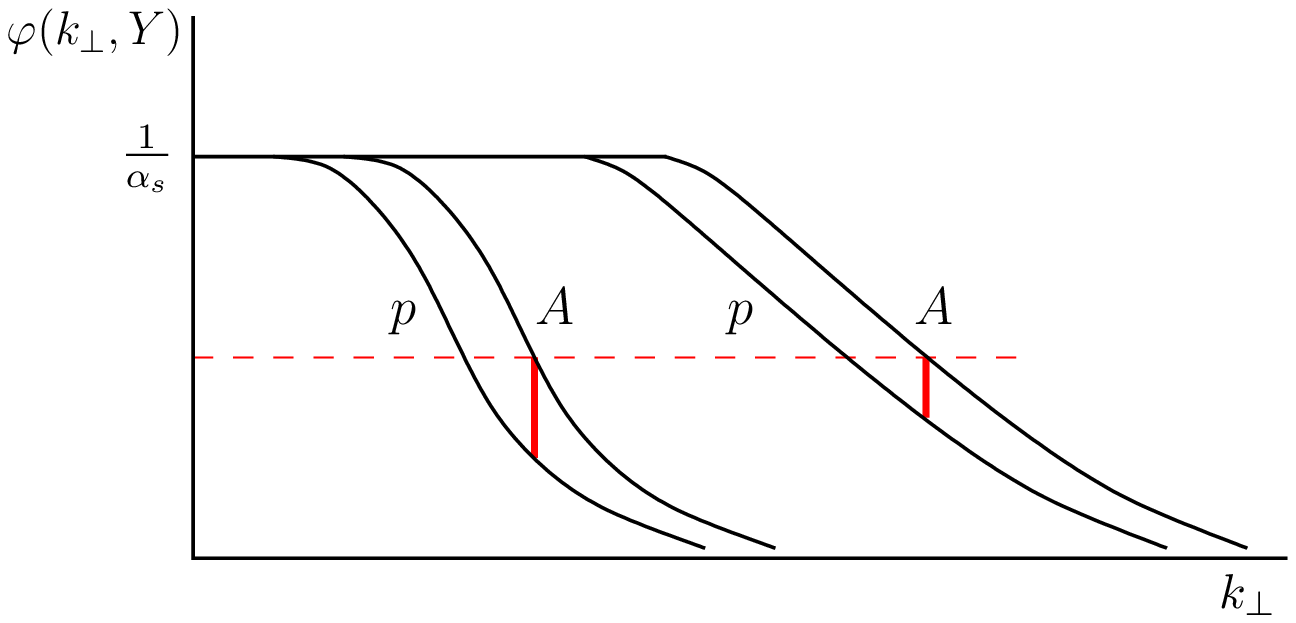, width=8.5cm}
\hspace*{0.3cm} (a) \hspace*{9cm} (b)
\end{center}
\caption{The qualitative behaviour of the unintegrated gluon distribution of a nucleus (A) and a proton (p)
  at two different rapidities in the geometric scaling regime (a) and diffusive
  scaling regime (b). }
\label{GS}
\end{figure}

In the geometric scaling regime the unintegrated gluon distributions $%
h_A(\rho-\langle \rho_s\rangle)$ and $\varphi_A(\rho-\langle\rho_s\rangle)$ show
the same behaviour (see Sect.~\ref{sec:gsr}). From Eqs.~(\ref{eq:h1}) and
(\ref{del_rho}) one finds for the ratio $R^{h,\varphi}_{pA}$ defined in Eq.~(\ref{R_pA})
\begin{eqnarray}
R^{h,\varphi}_{pA}(k_{\perp},Y,A)
    &\simeq& \frac{1}{A^{\frac{1}{3}}}
         \left[\frac{\langle Q_{s}(A,Y)\rangle^2}
         {\langle Q_{s}(p,Y)\rangle^2}\right]^{\gamma _{c}}\,
         \frac{\ln \left( \frac{k_{\perp }^{2}}
        {\langle Q_{s}(A,Y)\rangle^2}\right) +
        \frac{1}{\gamma _{c}}}{\ln \left(
        \frac{k_{\perp }^{2}}{\langle Q_{s}(p,Y)\rangle^2}\right) +
        \frac{1}{\gamma _{c}}}
\nonumber \\
&\simeq& \frac{1}{A^{\frac{1}{3}\left(1-\gamma _{c}\right) }}\
         \frac{\left[\alpha_s^2\,\ln(\alpha_s^2\,A^{1/3})
         \right]^{\gamma_c}}{\alpha_s^2 \ln(1/\alpha_s)}\
         \left( \frac{c^{h,\varphi}_A\, c\, x_{\perp}^{\prime\,2}\,Q_0^2}
         {c^{h,\varphi}_p }\right)^{\gamma_c}\
         \frac{\ln \left( \frac{k_{\perp}^{2}}
         {\langle Q_{s}(A,Y)\rangle^2}\right) +
         \frac{1}{\gamma _{c}}}{\ln \left(
         \frac{k_{\perp }^{2}}{\langle Q_{s}(A,Y) \rangle^2}\right) +
         \Delta \rho_s + \frac{1}{\gamma _{c}}}
\label{R_hd}
\end{eqnarray}
where we have used $h^{max}_A=h^{max}_p$ and $\varphi^{max}_A=\varphi^{max}_p$
and the average saturation momentum $\langle Q_s \rangle^2$ defined via $\langle
\rho_s \rangle \equiv \ln \langle Q_s \rangle^2/k_0^2$. For gluon momenta not
too far from $\langle Q_s(A,Y) \rangle$ the ratio
$R^{h,\varphi}_{\mbox{\scriptsize{pA}}}$ scales with $A$ like
$A^{1/3(\gamma_c-1)}$, rather than being equal to one. This is partial gluon
shadowing due to the anomalous behaviour of the unintegrated gluon distribution
given in Eq.~(\ref{h}) which stems from the BFKL evolution.  This result,
already derived in Refs.~\cite{Mueller:2003bz,Iancu:2004bx}, may explain why
particle production in heavy ion collisions scales, roughly, like
$N_{part}$~\cite{Kharzeev:2002pc}. Moreover
$R^{h,\varphi}_{\mbox{\scriptsize{pA}}}$ reduces to a $k_{\perp}^2$ and $Y$
independent
expression for $k^2_{\perp}$ much larger than $\langle Q_s(A,y)\rangle^2$, or $%
\ln(k^2_{\perp}/\langle Q_s(A,Y)\rangle^2 ) \gg \Delta \rho_s$. This comes from
the fact that the unintegrated gluon distribution of the nucleus and of the
proton preserves the shape with rising rapidity, yielding therefore a constant
value for their ratio. This behaviour is shown in Fig.~\ref{GS}(a).

\begin{figure}[htb]
\setlength{\unitlength}{1.cm}
\par
\begin{center}
\epsfig{file=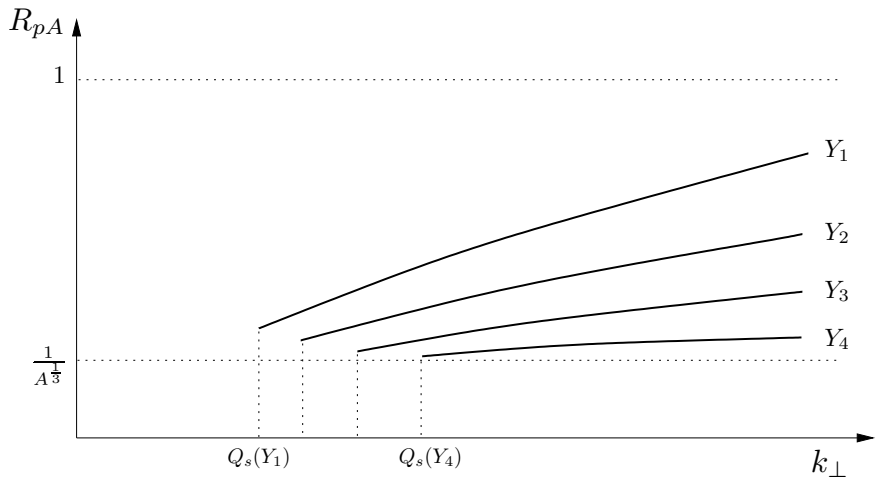, width=12cm}
\end{center}
\caption{The qualitative behaviour of the ratio $R_{pA}$ as a function of $k_{\perp}$ at four different
  rapidities, $Y_1 \leq Y_2 \leq Y_3 \leq Y_4$, in the diffusive scaling regime.
  $R_{pA}$ is always smaller than one for values of $k_{\perp}$ in the diffusive
  scaling regime.}
\label{R_DS}
\end{figure}

In the diffusive scaling regime the ratio $R_{pA}$ for the unintegrated
gluon distribution in Eq.~(\ref{eq:h_a_diff}) equals
\begin{eqnarray}
R^h_{pA}(k_{\perp},Y,A) &=& \frac{1}{A^{\frac{1}{3}}}\
\left[\frac{\langle Q_{s}(A,Y)\rangle^2}{\langle
Q_{s}(p,Y)\rangle^2}\right]^{\frac{\Delta \rho_s}{2\sigma^2}}\
\left[\frac{k_{\perp }^{2}}{\langle
Q_{s}(A,Y)\rangle^2}\right]^{
\frac{\Delta \rho_s}{\sigma^2}} \nonumber \\
&=&
\frac{1}{A^{\frac{1}{3}(1-\frac{\Delta\rho_s}{2\sigma^2})}} \
\left[\frac{\alpha_s^2\,\ln(\alpha_s^2\,A^{1/3})}
{(\alpha_s^2
  \ln(1/\alpha_s))^{1/\gamma_c}}\right]^{\frac{\Delta\rho_s}{2\sigma^2}} \
\left[\frac{c^{h}_A\,
c\,x_{\perp}^{\prime\,2}\,Q^2_0}{c^{h}_p }
         \right]^{\frac{\Delta\rho_s}{2\sigma^2}}\
\left[\frac{k_{\perp }^{2}}{\langle Q_{s}(A,Y)\rangle^2}\right]^{
\frac{\Delta \rho_s}{\sigma^2}}
\label{R_h_f}
\end{eqnarray}
while for the "Weizsacker-Williams" unintegrated gluon distribution in Eq.~(%
\ref{eq:phi_a_diff}) it is
\begin{eqnarray}
\!\!\!\!\!R^{\varphi}_{pA} &=& \frac{1}{A^{\frac{1}{3}}}\
\left[\frac{\langle Q_{s}(A,Y)\rangle^2}{\langle
Q_{s}(p,Y)\rangle^2}\right]^{\frac{\Delta \rho_s}{2\sigma^2}}\
\left[\frac{k_{\perp }^{2}}{\langle
Q_{s}(A,Y)\rangle^2}\right]^{ \frac{\Delta \rho_s}{\sigma^2}} \
\frac{\left[ \ln \left( \frac{k_{\perp }^{2}}{\langle
Q_{s}(A,Y)\rangle^2}\right)+\Delta\rho_s\right] ^{2}}{\left[ \ln
\left( \frac{k_{\perp }^{2}}{\langle
Q_{s}(A,Y)\rangle^2}\right)\right] ^{2}}
\nonumber \\
\!\!\!&=&
\frac{1}{A^{\frac{1}{3}(1-\frac{\Delta\rho_s}{2\sigma^2})}}
\left[\frac{\alpha_s^2\,\ln(\alpha_s^2\,A^{1/3})}
{(\alpha_s^2
  \ln(1/\alpha_s))^{1/\gamma_c}}\right]^{\frac{\Delta\rho_s}{2\sigma^2}}
\left[\frac{c^{\varphi}_A\, c
\,x_{\perp}^{\prime\,2}\,Q^2_0}{c^{\varphi}_p}
         \right]^{\frac{\Delta\rho_s}{2\sigma^2}}
\left[\frac{k_{\perp }^{2}}{\langle Q_{s}(A,Y)\rangle^2}\right]^{
\frac{\Delta \rho_s}{\sigma^2}}
\frac{\left[ \ln \left( \frac{k_{\perp }^{2}}{\langle
Q_{s}(A,Y)\rangle^2}\right)+\Delta\rho_s\right] ^{2}}{\left[ \ln \left(
\frac{k_{\perp }^{2}}{\langle Q_{s}(A,Y)\rangle^2}\right)\right] ^{2}} \ .
\label{R_v_f}
\end{eqnarray}
In the diffusive scaling regime $R^{h,\varphi}_{pA}$ becomes independent of
the definition of the gluon distribution and shows the universal
behaviour $A^{1/3(\frac{\Delta\rho_s}{2\sigma^2}-1)} \left
[k_{\perp}^2/\langle Q_s(A,Y)\rangle^2\right]^{\frac{\Delta\rho_s}{\sigma^2}}$ when $%
\ln(k^2_{\perp}/\langle Q_s(A,Y)\rangle^2) \gg \Delta\rho_s$. The above ratios in the
diffusive scaling regime show two different features as compared to the
ratio in the geometric scaling regime: (i) For $k^2_{\perp}$ close to $%
\langle Q_s(A,Y)\rangle^2$, the gluon shadowing characterized by
$A^{1/3(\frac{\Delta\rho_s}{2\sigma^2}-1)}$ is dominated by fluctuations, through
$\sigma^2(Y)$, and depends also on the difference $\Delta\rho_s$. The gluon shadowing
increases as the rapidity increases because of $\sigma^2 = D_{dc} \bar{\alpha}_s Y$. At asymptotic rapidity, where $\sigma^2 \to \infty$%
, one obtains \textit{total gluon shadowing}, $R^{h,\varphi}_{pA}=1/A^{1/3}$,
which means that the unintegrated gluon distribution of the nucleus and that of
the proton become the same in the diffusive scaling regime. The total gluon
shadowing is an effect of fluctuations at fixed couping since the fluctuations make the
unintegrated gluon distributions of the nucleus and of the proton flatter and
flatter and their ratio closer and closer to $1$ (at fixed $\Delta\rho_s$) with
rising rapidity, as shown in Fig.~\ref{GS}(b).  (ii) $R^{h,\varphi}_{pA}$ shows an increase with rising $%
k^2_{\perp}$ within the diffusive scaling region. Since the exponent
$\Delta\rho_s/\sigma^2$ decreases with rapidity, the slope of
$R^{h,\varphi}_{pA}$ as a function of $k_{\perp}^2$ becomes smaller with
increasing $Y$. Both features of $R_{pA}$ in the
diffusive scaling regime, the increasing gluon shadowing and the descreasing
$k_{\perp}$ dependence with rising rapidity, are shown in Fig.~\ref{R_DS}.  
Note also that in the common diffusive scaling regime, see (\ref{eq:dif_scal_reg}) and (\ref{eq:ods}), the ratio $R_{pA}$ is always smaller than one. (This
can be seen especially from Eq.(\ref{eq:A_R_DS}) where $\Delta\rho_s = \ln
A^{1/3}$ was used.)

Total gluon shadowing is not possible in the geometric scaling regime in the
fixed-coupling case since the shapes of the gluon distributions of the nucleus
and of the proton remain the same with increasing $Y$ giving for their ratio a
value unequal one (see Fig~\ref{GS}(a)). In the case of a running coupling, the
gluon shadowing increases with rising rapidity in the geometric scaling
regime~\cite{Iancu:2004bx}, as oppossed to the (roughly) fixed value (partial
shadowing) in the fixed-coupling case, and would lead to total gluon shadowing
at very high rapidities if fluctuations were absent. The combination of the
running of the coupling plus fluctuations would give total gluon shadowing at
lower rapidities as compared to the fixed coupling case presented here. An
extension of this work by the running coupling remains work for the future. 


%

\begin{acknowledgments}
  We would like to thank Al Mueller for valuable discussions. B. X. wishes to
  thank the Physics Department of the University of Bielefeld for hospitality
  during his visit when this work was being initialized. A. Sh. and M. K.
  acknowledge financial support by the Deutsche Forschungsgemeinschaft under
  contract Sh 92/2-1.

\end{acknowledgments}

\appendix

\section{McLerran-Venugopalan initial condition for the proton}
\label{MV_proton}
Also the proton can be described within the McLerran-Venugopalan model, as done
for the nucleus in Sect.~\ref{sec:init_cond}. The only change as compared to the
description of the nucleus is a different saturation momentum for the proton due
to the different initial conditions. It is reasonable to assume the following
relation between their saturation momenta
\begin{equation}
\langle Q_{s}(A,Y) \rangle^2 = A^{1/3}\ \langle Q_{s}(p,Y) \rangle^2
\label{eq:rel_A_p}
\end{equation}
or in terms of $\langle \rho
_{s}(A,Y)\rangle = \ln (\langle Q_{s}(A,Y)\rangle^2 /k_{0}^{2})$,
\begin{equation}
\langle \rho _{s}(A,Y)\rangle = \langle \rho _{s}(p,Y)\rangle + \ln A^{1/3}\ .
\label{mvv}
\end{equation}
This relation holds formally if one extrapolates the McLerran-Venugopalan model,
which was constructed for a large nucleus, down to $A=1$. So, to get the
expressions for the proton from the ones given for the nucleus in the previous
sections, one has to replace $\langle Q_{s}(A,Y) \rangle^2$ by $\langle
Q_{s}(p,Y) \rangle^2$, or, the difference $\Delta\rho_s$ in Eq.~(\ref{del_rho}) by
\begin{equation}
\Delta\rho^{MV}_s \equiv  \langle \rho _{s}(A,Y)\rangle -
\langle \rho _{s}(p,Y)\rangle = \ln A^\frac{1}{3} \ .
\end{equation}
With this difference, the ratio in the geometric scaling region, as compared to
(\ref{R_hd}), now reads
\begin{equation}
R^{h,\varphi}_{pA}(k_{\perp},Y,A) =
\frac{1}{A^{\frac{1}{3}\left(1- \gamma_{c}\right)}}\ \frac{\ln \left( \frac{k_{\perp
}^{2}}{\langle Q_{s}(A,Y)\rangle^2}\right) +\frac{1}{\gamma _{c}}}{\ln \left( \frac{%
k_{\perp }^{2}}{\langle Q_{s}(A,Y)\rangle^2}\right) + \ln A^{\frac{1}{3}} + \frac{1}{%
\gamma _{c}}} \
\end{equation}
and shows the same features (partial gluon shadowing and $k_{\perp}^2$ and $Y$
independence for $\ln k^2_{\perp}/\langle Q_s(A,Y) \rangle \gg \ln A^{1/3}$) as
the result in Eq.~(\ref{R_hd}). In the diffusive scaling regime, instead of
Eq.(\ref{R_h_f}), now one obtains
\begin{equation}
R^h_{pA} = \frac{1}{A^{\frac{1}{3}\left(1-\frac{\ln A^{1/3}}{2\sigma^{2}}\right)}} \
      \left[\frac{k_{\perp }^{2}}{\langle Q_{s}(A,Y)\rangle^2}\right]^{\frac{\ln
      A^{1/3}}{\sigma^{2}}} \ ,
\label{eq:A_R_DS}
\end{equation}
and instead of Eq.~(\ref{R_v_f})
\begin{equation}
R^{\varphi}_{pA} = \frac{1}{A^{\frac{1}{3}
                \left(1- \frac{\ln A^{1/3}}{2\sigma^{2}}\right)}}
  \left[ \frac{k_{\perp }^{2}}{\langle Q_{s}(A,Y)\rangle^2}\right]^{\frac{\ln
A^{1/3}}{\sigma ^{2}}}\frac{\left[ \ln \left( \frac{k_{\perp }^{2}}{%
\langle Q_{s}(A,Y)\rangle^2}\right)+\ln A^{\frac{1}{3}}\right] ^{2}}{\left[ \ln \left(
\frac{k_{\perp }^{2}}{\langle Q_{s}(A,Y)\rangle^2}\right)\right] ^{2}} \ ,
\end{equation}%
which show the same universal features for the ratio $R^{h,\varphi}_{pA}$, total
shadowing at $Y \to \infty$ and a decreasing $k^2_{\perp}$ dependence with rising
$Y$, as discussed in the previous section.

\section{Gluon distribution in the saturation region}
\label{sec:ap}
We calculate the gluon distribution in the saturation region, where
$k_{\perp }^{2} < Q_{s}^{2}$, by solving the Kovchegov
equation~\cite{Kovchegov:1999yj+X} at high rapidities. The Kovchegov equation in
momentum space, at a fixed impact parameter, reads~\cite{Kovchegov:1999yj+X}
\begin{equation}
\frac{\partial \,\phi \left( \rho ,Y\right) }{\overline{\alpha}_{s}\partial
Y}=\chi \left( -\partial _{\rho }\right) \phi \left( \rho ,Y\right) -\phi
^{2}\left( \rho ,Y\right) \ ,  \label{kov}
\end{equation}
where $\rho =\ln \frac{k_{\perp }^{2}}{k_{0}^{2}}$ and $\phi \left( \rho
  ,Y\right) = \frac{\left( 2\pi \right)
  ^{2}\alpha_{s}}{N_{c}}\,\varphi_{A}\left( k_{\perp },Y\right)$ the
Weizsacker-Williams gluon distribution defined in Eq.~(\ref{sa}) (apart from the
prefactor).  It is convenient to look for a solution to the Kovchegov equation
in the saturation regime in the form
\begin{eqnarray}
\phi \left( \rho ,Y\right) &=&\int_{c-i\infty }^{c+i\infty }\frac{d\gamma }{%
2\pi i}\phi \left( \gamma \right) \exp \left[ -\gamma \left( \rho -v%
\bar{\alpha}_{s}Y\right) \right] , \\
&=&\int_{c-i\infty }^{c+i\infty }\frac{d\gamma }{2\pi i}\phi \left( \gamma
\right) \exp \left[ -\gamma \ln \left( \frac{k_{\perp }^{2}}{Q_s^{2}(Y)}\right) \right]
\end{eqnarray}
with $Q_{s}^{2}(Y)  = k_{0}^{2}\,\exp \left[ \bar{\alpha}_{s}\,v\,Y\right]$. In the saturation regime, $\ln \left( \frac{k_{\perp
}^{2}}{Q_{s}^{2}(Y)}\right) <0$, the contour
integral has to be closed to the left and form a counterclockwise path
in order to enclose all the poles of $\phi \left( \gamma \right) $. The first and most important pole of $\phi \left(
\gamma \right) $ is at $\gamma =0$. Using the Laurent expansion of $%
\phi \left( \gamma \right) $ at $\gamma =0$, which is $\phi \left( \gamma \right)
=\sum\limits_{n=-\infty }^{\infty }a_{n}\gamma ^{n}$, in Eq.(\ref{kov}), one can immediately spot that there must be a cutoff
of $n$ at $n=-2$. Thus, $\phi \left( \gamma \right) $ can be written as
\begin{equation}
\phi \left( \gamma \right) =\frac{a_1}{\gamma ^{2}}+\frac{a_2}{\gamma }%
+a_3+f\left( \gamma \right) \
\end{equation}
with $f(0)=0$. The function $f\left( \gamma \right) $ would have some poles at $\gamma
=-\gamma _{i}$ where $\gamma _{i}>0$, which however do generate
small contributions like $\left( \frac{k_{\perp }^{2}}{Q_{s}^{2}\left( Y\right) }\right) ^{\gamma _{i}}$ to $\phi \left( \rho
,Y\right) $ in the saturation region. Therefore the most important contributions are from $\phi
_{0}\left( \gamma \right) \equiv \phi(\gamma) - f(\gamma)$.

It is now straightforward to see that
\begin{eqnarray}
\phi _{0}\left( \rho ,Y\right) &=&a_1\,\ln \frac{Q_{s}^{2}\left(
Y\right) }{k_{\perp }^{2}}+a_2, \\
\frac{\partial \,\phi _{0}\left( \rho ,Y\right) }{\overline{\alpha}_{s}%
\partial Y} &=&a_1\,v, \\
\chi \left( -\partial _{\rho }\right) \phi _{0}\left( \rho ,Y\right) &=&%
\frac{1}{2}\,a_1\,\left( \ln \frac{Q_{s}^{2}\left( Y\right) }{k_{\perp
}^{2}}\right) ^{2}+a_2\ln \frac{Q_{s}^{2}\left( Y\right) }{k_{\perp
}^{2}}+a_3+\sum\limits_{n=1}^{\infty }\left( \frac{a_1}{n^{2}}-\frac{a_2}{n}%
+a_3\right) \left( \frac{k_{\perp }^{2}}{Q_{s}^{2}\left( Y\right) }%
\right)^{n}
\label{bdc}
\end{eqnarray}
where the last term in (\ref{bdc}) comes from the residues of $\chi \left( \gamma \right) $ at $%
\gamma =-n$  with $n=1,2,...$. Up to some small correction of order $\frac{%
k_{\perp }^{2}}{Q_{s}^{2}\left( Y\right) }$, Eq.(\ref{kov}) is
solved by $\phi _{0}\left( \rho ,Y\right) =a_1\,\ln \frac{Q_{s}^{2}\left( Y\right) }{k_{\perp }^{2}}+a_2$ with $a_1=\frac{1}{2}$ and $\frac{%
1}{2}v=a_3-a_2^{2}$. Moreover one can set $a_2=0$ and get $a_3=\frac{1}{2}v$.
Finally, one obtains the approximate solution to the Kovchegov equation,  $%
\phi \left( \rho ,Y\right) =\frac{1}{2}\ln \frac{Q_{s}^{2}\left(
Y\right) }{k_{\perp }^{2}}+{\cal{O}}\left( \frac{k_{\perp }^{2}}{Q_{s}^{2}\left(
  Y\right) }\right) $, and the gluon distribution as defined in Eq.~(\ref{sa})
takes the form in the saturation regime as cited in Eq.(\ref{varphi_k})
\begin{equation}
\varphi _{A}\left( k_{\perp },Y\right) =\frac{N_{c}}{2\left( 2\pi \right)
^{2}\alpha _{s}}\left[ \ln \frac{Q_{s}^{2}\left( Y\right) }{%
k_{\perp }^{2}}+{\cal{O}}\left( \frac{k_{\perp }^{2}}{Q_{s}^{2}\left(
Y\right) }\right) \right] \ .
\label{var_keq}
\end{equation}
The gluon distribution as defined in Eq.(\ref{k}) reads in the saturation
regime
\begin{equation}
h_{A}\left( k_{\perp },Y\right) \propto \frac{N_{c}}{2\left( 2\pi
\right) ^{2}\alpha _{s}}\ \frac{k_{\perp }^{2}}{Q_{s}^{2}\left(
Y\right) },
\end{equation}
and is obtained from Eq.~(\ref{var_keq}), the higher order correction term, by
using the relation in Eq.~(\ref{eq:re;_h_v}).


\end{document}